\documentclass[conference]{IEEEtran}
\IEEEoverridecommandlockouts
\usepackage{cite}
\usepackage{mathrsfs}
\usepackage{amsmath,amssymb,amsfonts}
\usepackage{algorithmic}
\usepackage{graphicx}
\usepackage{textcomp}
\usepackage{xcolor}
\usepackage{multirow}
\usepackage{booktabs}
\def\BibTeX{{\rm B\kern-.05em{\sc i\kern-.025em b}\kern-.08em
    T\kern-.1667em\lower.7ex\hbox{E}\kern-.125emX}}
\begin{document}

\title{HingeNet: A Harmonic-Aware Fine-Tuning Approach for Beat Tracking}

\author{
\IEEEauthorblockN{
Ganghui Ru\textsuperscript{1, $\star$}
Jieying Wang\textsuperscript{2, $\star$},
Jiahao Zhao\textsuperscript{3},
Yulun Wu\textsuperscript{1},
Yi Yu\textsuperscript{4},
Nannan Jiang\textsuperscript{2, $\dag$},
Wei Wang\textsuperscript{2, $\dag$},
and Wei Li\textsuperscript{1, 5, $\dag$} 
}

\IEEEauthorblockA{
\textsuperscript{1}School of Computer Science, Fudan University, Shanghai, China \\
\textsuperscript{2}Naval Medical Center, PLA, China \\
\textsuperscript{3}Graduate School of Informatics, Kyoto University, Kyoto, Japan \\
\textsuperscript{4}Graduate School of Advanced Science and Engineering, Hiroshima University, Hiroshima, Japan \\
\textsuperscript{5}Shanghai Key Laboratory of Intelligent Information Processing, Fudan University, Shanghai, China
}

\IEEEauthorblockA{
ghru23@m.fudan.edu.cn, 
\{20110240018, wwang\_fd, weili-fudan\}@fudan.edu.cn, 
\{wanwan0426, jiangnannannavy\}@163.com, 
 \\
zhao.jiahao.56h@st.kyoto-u.ac.jp, 
yiyu@hiroshima-u.ac.jp 
}
}

\maketitle
\footnotetext{$\star$These authors contributed equally to this work}
\footnotetext{$\dag$Corresponding author}

\begin{abstract}
Fine-tuning pre-trained foundation models has made significant progress in music information retrieval. 
However, applying these models to beat tracking tasks remains unexplored as the limited annotated data renders conventional fine-tuning methods ineffective. 
To address this challenge, we propose HingeNet, a novel and general parameter-efficient fine-tuning method specifically designed for beat tracking tasks. 
HingeNet is a lightweight and separable network, visually resembling a hinge, designed to tightly interface with pre-trained foundation models by using their intermediate feature representations as input. 
This unique architecture grants HingeNet broad generalizability, enabling effective integration with various pre-trained foundation models. 
Furthermore, considering the significance of harmonics in beat tracking, we introduce harmonic-aware mechanism during the fine-tuning process to better capture and emphasize the harmonic structures in musical signals.
Experiments on benchmark datasets demonstrate that HingeNet achieves state-of-the-art performance in beat and downbeat tracking.
\end{abstract}


\section{Introduction}
\label{sec:intro}

As digital music continues to evolve, accurately identifying rhythmic patterns within compositions has become a key focus in Music Information Retrieval (MIR). Beat tracking is a critical component of this field due to its foundational role in defining the temporal framework for analyzing musical content. The goal of beat tracking is to detect the temporal positions of beats within a music signal. Accurate beat tracking is essential for advanced tasks such as music transcription \cite{transcription1, transcription2} and music structure analysis \cite{Structure_Analysis1, Structure_Analysis2}, among others \cite{ru2023improving, duan2023melody}. However, beat tracking remains a challenging task, primarily due to the complex and varied nature of rhythmic structures in different musical genres. 

\begin{figure}[ht]
	\centering
	\includegraphics[width=0.9\linewidth]{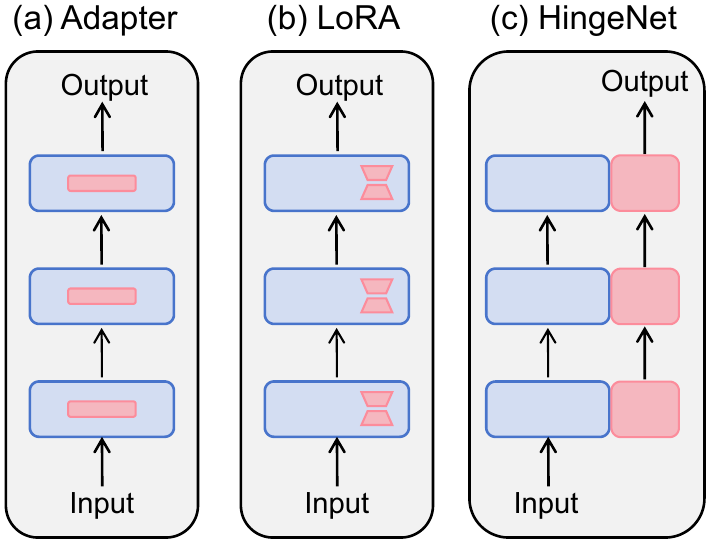}	
	\caption{Comparison between the proposed HingeNet and traditional fine-tuning methods, such as Adapter and LoRA. The blue blocks represent the frozen encoder of the pre-trained music foundation model, while the red blocks indicate the trainable fine-tuning modules.
 }
	\label{figure1}
\end{figure}

With the rise of deep learning, methods have significantly evolved. Recurrent Neural Networks (RNNs) \cite{rnn1, rnn2} are used to capture temporal dependencies, while Convolutional Neural Networks (CNNs) \cite{cnn1, cnn2} are employed to learn local patterns from spectrograms, improving beat detection. Temporal Convolutional Networks (TCNs) \cite{tcn1, bock2020} are designed to provide better handling of long-range dependencies compared to RNNs. More recently, Transformer-based models \cite{tftrans, beattrans, lhtrans} have emerged, leveraging self-attention to capture long-range rhythmic patterns, achieving state-of-the-art results in beat tracking.

While these methods have led to  some performance improvements, they inevitably reach a performance plateau due to limited representation capacity.
Fine-tuning pre-trained foundation models allows for the transfer of their extensive knowledge to specific downstream tasks, which has already proven effective across various domains \cite{fine2}.
Therefore, extending this strategy to beat tracking tasks to overcome current performance limitations is a natural and promising approach. 
As shown in Fig.1(a) and 1(b), traditional fine-tuning methods, such as Adapter\cite{adapter} and LoRA \cite{lora}, typically insert trainable fine-tuning modules into frozen foundation models, which alters the model’s architecture and representation space.
Previous research \cite{lin2024lora} has shown that these methods are less effective in tasks with limited annotated data, as they can lead to overfitting and disrupt the generalizability of the pre-trained model’s features.


To address these issues, we propose a parameter-efficient fine-tuning method that visually resembles a hinge, which we therefore name HingeNet.  As shown in Fig.1(c), HingeNet is designed as an independent architecture. This unique architecture tightly integrates with various pre-trained foundation models by taking intermediate feature representations as input, ensuring that the fine-tuning fully leverages robust feature representations learned by the pre-trained model. 
Furthermore, according to music theory, harmonic shifts are more likely to occur at beat positions rather than non-beat positions. To capture these shifts effectively, we introduce harmonic-aware mechanism within HingeNet, which enhances the model's ability to distinguish beat-related features, further improving beat and downbeat tracking accuracy. In summary, our contributions are as follows:
\begin{itemize}
  \item We propose HingeNet, a lightweight fine-tuning method specifically designed for beat tracking tasks.  Its separable architecture enables effective integration with various pre-trained foundation models, ensuring strong generalization and superior performance.
  \item Considering the significance of harmonics, we introduce harmonic-aware mechanism designed to capture harmonic shifts at beat positions and emphasize harmonic structures, thereby enhancing the accuracy of beat and downbeat tracking.
  \item Extensive experiments on multiple benchmark datasets demonstrate that HingeNet achieves state-of-the-art performance in both beat and downbeat tracking.
\end{itemize}

\begin{figure*}[ht]
	\centering
	\includegraphics[width=0.95\linewidth]{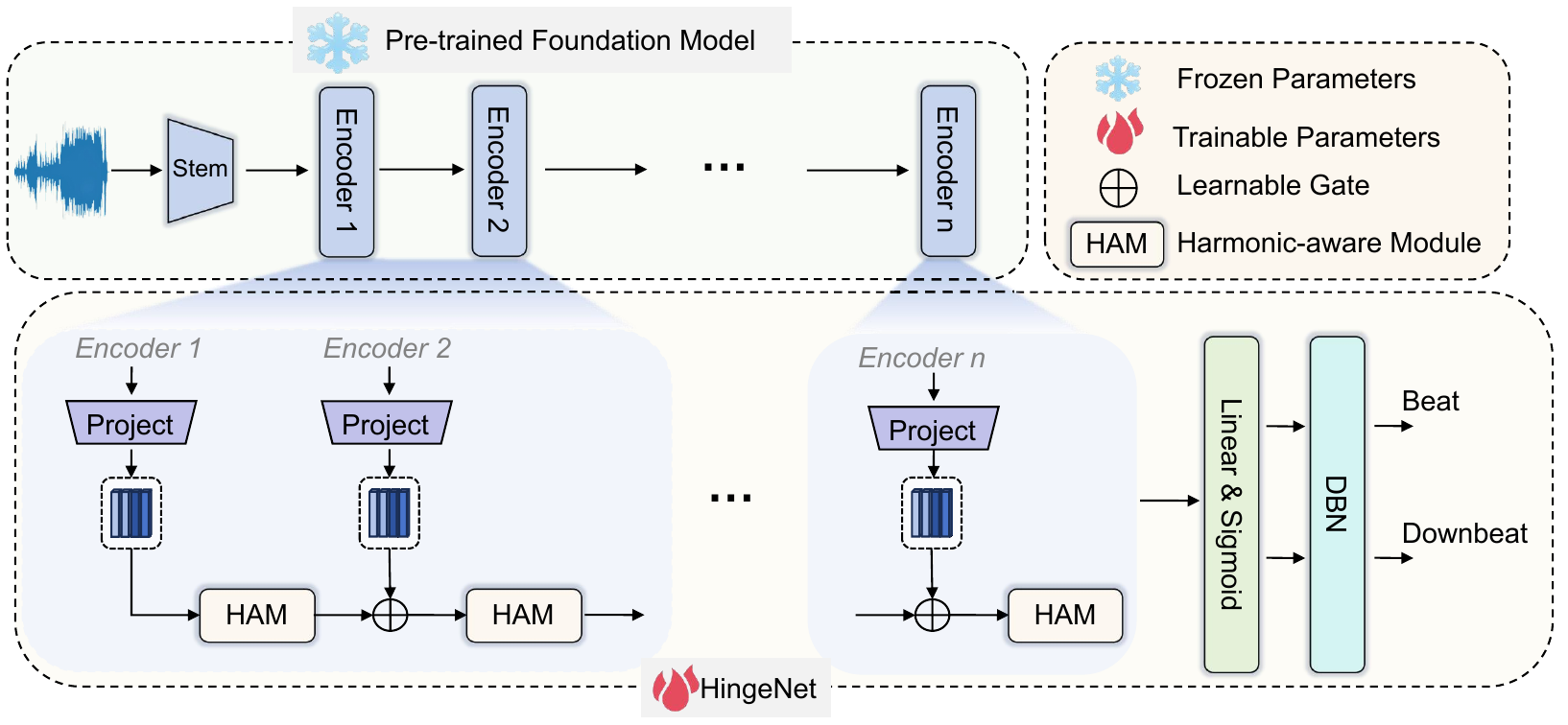}	
	\caption{Overview of our proposed model, consisting of two parts: the pre-trained foundation model and the HingeNet.}
	\label{figure1}
\end{figure*}

\section{Related Work}
\label{sec:format}
\subsection{Beat and Downbeat Tracking}
Early methods, including RNNs \cite{rnn1, rnn2}, CNNs \cite{cnn1, ubeat}, and TCNs \cite{tcn1, bock2020} , were used to capture temporal and spectral patterns. However, these models struggled with long-range dependencies and complex rhythmic structures, especially in music with varying tempos.

More recently, Huang \textit{et al.} \cite{tftrans} introduced Transformer-based models for beat tracking, significantly improving the accuracy of the model. Following this, Zhao \textit{et al.} \cite{beattrans} and Cheng \textit{et al.} \cite{lhtrans} proposed further improvements to Transformer-based models, optimizing them for better handling of complex rhythms and varying tempos. In addition, Heydari \textit{et al.} \cite{singingbeattracking} explored using pre-trained self-supervised speech representation models as feature extractors for singing beat tracking. Desblancs \textit{et al.} \cite{zeronotesamba} investigated unsupervised beat tracking methods, aiming to eliminate the need for labeled data in training. Meanwhile, Chiu \textit{et al.} \cite{chiu2023local} and Foscarin \textit{et al.} \cite{beatthis} explored improvements to, and even removal of, the DBN post-processing step, seeking more efficient alternatives for refining beat predictions. Despite these innovations, challenges remain in handling diverse musical styles and achieving accurate beat tracking in complex audio sources.

The emergence of music foundation models has brought new hope in addressing these challenges. Pre-training  on large-scale music data endows these models with robust semantic representations and strong generalization capabilities. Fine-tuning for specific downstream tasks can further enhance the performance of these models. Motivated by the unique characteristics of beat tracking, we propose HingeNet, a novel harmonic-aware fine-tuning method. This approach significantly improves the performance of foundation models in beat tracking tasks by focusing on harmonic shifts, enhancing both accuracy and efficiency.

\subsection{Foundation Models in MIR}
The exploration of foundation models within MIR is still in its early stages, but some groundbreaking work has already been done. Music2Vce \cite{map} employs a student-teacher framework to enhance its performance in music understanding tasks. In this setup, the student network learns from the output of the teacher network, allowing it to effectively capture complex musical patterns and improve generalization across diverse musical genres. Building on this approach, MERT \cite{mert} utilizes a Residual Vector Quantization - Variational AutoEncoder (RVQ-VAE) as the acoustic teacher and a Constant-Q Transform (CQT) as the musical teacher to guide the model in jointly learning both acoustic and musical knowledge. This combination enables MERT to effectively capture the tonal and pitched characteristics inherent in music, making it well-suited for local frame-level sequence labeling tasks. 

Additionally, MusicFM \cite{musicfm} draws inspiration from the speech recognition model BEST-RQ \cite{bestrq}, using a random projection quantizer during the tokenization phase. This approach significantly improves the performance of frame-level classification tasks that require long-term contextual understanding. To fully harness the potential of foundation models for beat tracking tasks, we propose a novel harmonic-aware fine-tuning method specifically designed for this purpose.

\section{Method}
\label{sec:format}
\subsection{The Overall Architecture of HingeNet}
As shown in Fig.2, our proposed model consists of two parts: the upper part illustrates the pre-trained foundation model with frozen parameters, where intermediate feature representations are extracted from each encoder layer. 
The lower part shows HingeNet, a harmonic-aware fine-tuning network specifically designed for beat tracking and adaptable to various pre-trained foundation models. 
Specifically, HingeNet utilizes the intermediate feature representations from each encoder layer of the foundation model as inputs, enabling it to make predictions without altering the architecture of the foundation model.
Additionally, HingeNet integrates harmonic-aware modules specifically designed for beat tracking, which directly address the unique challenges of this task by refining and emphasizing harmonic structures in musical signals. 
This focuses on both the complete preservation of the pre-trained model and task-specific adaptation distinguishes HingeNet from other similar approaches.

Music foundation models typically consist of a stem module $\mathcal{S}$, which is usually composed of convolutional layers to convert input audio into time-frequency features. These features are then passed through $N$ Transformer encoders, denoted as $\{\mathcal{F}_i\}^N_{i=1}$, where each encoder captures different levels of musical semantics in the audio signal. After each encoder, we attach a corresponding HingeNet core layer, represented as $\{\mathcal{P}_i, \mathcal{H}_i\}^N_{i=1}$. Here, $\mathcal{P}$ represents the projection layer, which reduces the dimensionality of the output features to remove redundant information, and $\mathcal{H}$ denotes the harmonic-aware module, which captures harmonic shifts in the music signal and improves the model's ability to track beats and downbeats.

Given input audio x, we obtain its corresponding semantic feature representation as:
\begin{equation}
\begin{aligned}
h_0^\mathcal{F} = \mathcal{S}(x) ;   h_i^\mathcal{F} = \mathcal{F}_i(h_{i-1}^\mathcal{F})
\end{aligned}
\end{equation}

For the $i$-th HingeNet core layer, we first feed the $i$-th encoder's output $h_i^\mathcal{F} \in \mathbb{R}^{b*h*t}$ into $i$-th
projection layer $\mathcal{P}_i$ to obtain the projected features $h_i^\mathcal{P} \in \mathbb{R}^{b*\frac{h}{r}*t}$:
\begin{equation}
\begin{aligned}
h_i^\mathcal{P} = \mathcal{P}_i(h_i^\mathcal{F})
\end{aligned}
\end{equation}

Next, the projected features $h_i^\mathcal{P}$ are fused with the output of the previous HingeNet core layer $h_{i-1} \in \mathbb{R}^{b*\frac{h}{r}*t}$ through a learnable gating mechanism, producing the input for the harmonic-aware module:
\begin{equation}
\begin{aligned}
\tilde{h}_i = \mu_i*h_{i-1} + (1-\mu_i)*h_i^\mathcal{P}, \quad \forall i \in \{2, \dots, N\}
\end{aligned}
\end{equation}
where $\mu_i = sigmoid (\alpha_i)$, with $\alpha_i$ as a learnable scalar initialized to zero. We also experimented with other fusion methods, such as element-wise addition and cross-attention, but found that the current design works the best. For the first layer ($i$ = 1), there is no prior core layer output, so $\tilde{h}_i$ is directly set to the projected features: $\tilde{h}_1 = h_1^\mathcal{P}$.

The fused features $\tilde{h}_i$ are processed through a harmonic-aware module, consisting of $M$ parallel 1D convolutional layers with different dilation rates.  The resulting feature are then concatenated and  passed through an MLP layer, which maps them back to the original dimensional space,  yielding the output of the $i$-th HingeNet core layer, $h_i \in \mathbb{R}^{b*\frac{h}{r}*t}$:

\begin{equation}
\begin{aligned}
h_i = MLP(Concat[\mathcal{H}_i^1(\tilde{h}_i), \mathcal{H}_i^2(\tilde{h}_i)..., \mathcal{H}_i^M(\tilde{h}_i)]))
\end{aligned}
\end{equation}


Finally, the last features $h_{N}$ pass through a linear layer with sigmoid activation, followed by DBN post-processing \cite{dbn}, to produce the final beat and downbeat tracking results.
\subsection{Lightweight HingeNet Core Layer}
The HingeNet core layer is designed to efficiently integrate harmonic-aware information while maintaining a lightweight and flexible structure. The core idea behind this design is to retain essential musical patterns while reducing unnecessary parameterization to minimize the reliance on large amounts of annotated data.  The layer consists of three main components: the projection layer, the learnable gating mechanism, and the harmonic-aware module.

\textbf{Projection Layer.} Music signals contain rich semantic information, some of which help identify the beat, such as harmonics. However, other elements may be unrelated or even introduce interference, making beat detection more difficult. The projection layer simplifies this complexity by introducing a projection factor $r$ (e.g. $r$ = 2, 4, 6, 8), which reduces the dimensionality of the input features to $\tfrac{1}{r}$. This reduction is particularly crucial for beat tracking tasks with limited annotated data, as it helps control the number of trainable parameters and enhances the model's scalability.

\textbf{Learnable Gating Mechanism.} The learnable gating mechanism introduces dynamic control over the integration of the outputs from the other two components: the projection layer and the previous harmonic-aware module.  Initially, the gating parameter assigns equal weight to both components, ensuring that the contributions are balanced. However, as training progresses, the gating parameter adapts, allowing the model to dynamically adjust the relative importance of each component based on the task and data characteristics. This flexibility enables the model to focus more on harmonic features or projection features as needed, optimizing performance for downstream beat and downbeat tracking tasks.

\textbf{Harmonic-Aware Module.}
Harmonics play a crucial role in music perception, as harmonic shifts are more likely to occur at beat positions than at non-beat positions, providing valuable cues for beat detection. In addition, harmonics follow predictable patterns in the time-frequency representation of music, which can be exploited to capture these harmonic shifts more effectively. For example, in the CQT spectrum, adjacent harmonics of a fundamental frequency $f_0$ maintain a constant frequency interval when the $Q$ is appropriately chosen. The interval $d_k$ between adjacent harmonics in the harmonic series can be calculated using:
\begin{equation}
\begin{aligned}
d_k &= \log_{2^{1/Q}}(f_0 \cdot (k + 1)) - \log_{2^{1/Q}}(f_0 \cdot k), \\
    &= Q \cdot \log_2\left(\frac{k + 1}{k}\right)
\end{aligned}
\end{equation}
Where $Q$ indicates the number of bins per octave, $d_k$ denotes the interval of adjacent harmonics, $k$ is the serial number of the harmonic series. When $Q$ = 12 and the harmonic series $N$ is 5, the intervals between harmonics become constant and can be rounded to integer values as follows: [12, 7, 5, 4]. 
By using multiple parallel 1D convolutions with carefully selected dilation rates that correspond to the harmonic intervals, the harmonic-aware module effectively captures harmonic patterns and detects beat-related shifts. We describe in Equation 4 how these features are further processed to obtain the final output.

\section{experiments}
\subsection{Datasets and Metrics}
Following previous work \cite{ubeat, lhtrans}, HingeNet is trained and evaluated on seven standard music datasets.
Specifically, the \textit{Beatles} \cite{beatles}, \textit{RWC Popular} \cite{rwc_popular}, and \textit{Harmonix} \cite{hainsworth} datasets are used exclusively for training.
The \textit{Ballroom} \cite{ballroom2}, \textit{Hainsworth} \cite{harmonix}, and \textit{SMC} \cite{smc} datasets are used for both training and testing using an 8-fold cross-validation approach. \textit{GTZAN} \cite{gtzan2} dataset is reserved solely for testing, serving as an independent benchmark to evaluate the model's generalization ability. These datasets cover diverse musical styles and genres, offering a robust  benchmark for beat and downbeat tracking.

To evaluate beat and downbeat tracking performance, we adopt three widely-used metrics: F1-measure, CMLt(Correct Metric Level), and AMLt (allowing for off-beat or double or half tempo), with a tolerance window of 70 ms \cite{beatles}. The latter two metrics primarily evaluate the proportion of correctly predicted beat sequences matching the ground truth, highlighting the model's ability to consistently predict rhythmic patterns.

\subsection{Experimental Setup}
Our model is based on a multi-task learning framework, where beat and downbeat tracking are predicted simultaneously. Binary cross-entropy loss is used to supervise the training process. Following \cite{bock2020}, we apply the same label broadening technique to annotations. Specifically, frames adjacent to the annotated beat frames (±2 frames) are also marked as beats, but with lower weights of 0.5 and 0.25, respectively.

During training, we use time stretching without altering pitch as the data augmentation technique to enhance the robustness and generalization of our model. The validation and test data are left untouched. We train our model using the Adam optimizer with a learning rate of 1e-3 and a batch size of 16. The training is stopped when the validation loss does not decrease for 20 epochs. The total number of trainable parameters in our model is 4.74M, which constitutes only 1.4\% of the total parameters in the pre-trained model.

\subsection{Results and Analysis}
We compare our model with several state-of-the-art (SOTA) models, including Beat trans \cite{beattrans}, LH trans \cite{lhtrans}, and Beat This \cite{beatthis}. Additionally, we establish two baseline models based on pre-trained music foundation models: MERT \cite{mert} and MusicFM \cite{musicfm}, which only connect a beat and downbeat classifier along with DBN on top of the frozen pre-trained models. 
To ensure a fair comparison, we reproduced MERT's results on the \textit{GTZAN} dataset, as the originally reported results did not adhere to the standard conventions of the beat tracking community. Likewise, for the Beat This model, which was trained using additional data, we report its results on the \textit{GTZAN} dataset under the standard training setup.

\begin{table*}[h]
  \centering
  \caption{Comparison with other state-of-the-art beat tracking models and two baseline models on the \textit{GTZAN} dataset.}
  \fontsize{5}{5}\selectfont
  \resizebox{\textwidth}{!}{
  
    \begin{tabular}{cccccccc}
    \toprule
          &       & \multicolumn{3}{c}{\textbf{Beat Accuracy}} & \multicolumn{3}{c}{\textbf{Downbeat Accuracy}} \\
    \cmidrule{3-8}    
    \textbf{Dataset} & \textbf{Model} & \textbf{F-Measure} & \textbf{CMLt} & \textbf{AMLt} & \textbf{F-Measure} & \textbf{CMLt} & \textbf{AMLt} \\
    \midrule
    \multirow{7}{*}{GTZAN} 
          & Beat trans \cite{beattrans} & 88.5 & 80.0 & 92.2 & 71.4 & 66.5 & 84.4 \\
          & LH trans \cite{lhtrans} & 88.4 & 80.8 & 94.0 & - & - & - \\
          & Beat This \cite{beatthis} & 88.9 & 79.9 & 89.4 & 75.5 & 60.8 & 75.5 \\
          & MERT \cite{mert} & 87.3 & 78.4 & 90.7 &  74.8 & 69.3 & 86.1 \\
          & MusicFM \cite{musicfm} & 86.1 & - & - & 78.5 & - & - \\
          \cmidrule{2-8}
          & MERT+HingeNet  & \textbf{89.7} & \textbf{81.4} & \textbf{94.3} & 77.4 & 71.6 & 87.2 \\
          & MusicFM+HingeNet  & 89.2 & 80.9 & 93.7 & \textbf{79.8} & \textbf{73.2} & \textbf{89.5} \\
    \bottomrule
    \end{tabular}%
    }
  \label{testing_results}%
\end{table*}%

\begin{table}[ht]
    \centering
    \caption{Comparison with other state-of-the-art beat tracking models on datasets used in an 8-fold cross-validation setup.}
    \label{table:beat_accuracy}
    \begin{tabular}{cccc}
        \toprule
        \textbf{Dataset} & \textbf{Model} & \textbf{Beat F1} & \textbf{Downbeat F1} \\
        \midrule
        \multirow{6}{*}{Ballroom}
        & Beat trans \cite{beattrans} & 96.8 & 94.1  \\
        & LH trans \cite{lhtrans} & 95.0 & - \\
        & MERT \cite{mert}  & 95.7 &  93.2 \\
        & MusicFM \cite{musicfm} & 95.1 & 94.3  \\
        & MERT+HingeNet  & \textbf{97.3} & 94.5  \\
        & MusicFM+HingeNet  & 97.0 & \textbf{95.8}  \\
        \midrule
        \multirow{6}{*}{Hainsworth}
        & Beat trans \cite{beattrans} & 90.2 & 74.8  \\
        & LH trans \cite{lhtrans} & 87.0 & - \\
        & MERT \cite{mert}  & 89.6 & 74.5  \\
        & MusicFM \cite{musicfm} & 89.2 & 75.7  \\
        & MERT+HingeNet  & \textbf{91.4} & 76.7  \\
        & MusicFM+HingeNet  & 90.8 & \textbf{78.3}  \\
        \midrule
        \multirow{6}{*}{SMC}
        & Beat trans \cite{beattrans} & 59.6 & -  \\
        & LH trans \cite{lhtrans} & 55.4 & - \\
        & MERT \cite{mert}  & 60.1 & -  \\
        & MusicFM \cite{musicfm} & 59.2 &  - \\
        & MERT+HingeNet  & \textbf{61.7} & -  \\
        & MusicFM+HingeNet  & 60.6 & -  \\
        \bottomrule
    \end{tabular}
    \label{other_dataset}%
\end{table}

\begin{table}[ht]
    \centering
    \caption{Ablation study on beat tracking with different fine-tuning method on the GTZAN dataset.}
    \label{ab_finetune}
    \begin{tabular}{cccc}
        \toprule
        \textbf{Model} & \textbf{fine-tuning method} & \textbf{Beat F1} & \textbf{Downbeat F1} \\
        \midrule
        \multirow{3}{*}{MERT}
        & Adapter & 73.4 & 64.9  \\
        & LoRA & 81.2 & 69.8 \\
        & HingeNet  & 89.7 & 77.4  \\
        \midrule
        \multirow{3}{*}{MusicFM}
        & Adapter & 77.3 & 65.5  \\
        & LoRA  & 80.6 & 68.7 \\
        & HingeNet  & 89.2 & 79.8  \\
        \bottomrule
    \end{tabular}
\end{table}

Table \ref{testing_results} shows the results of HingeNet compared to several SOTA models and two baseline models on the \textit{GTZAN} dataset. 

\textbf{Stable Improvement with HingeNet.} Compared to the baseline models, fine-tuning with HingeNet consistently results in significant improvements across all metrics. This demonstrates the effectiveness of our harmonic-aware fine-tuning method, positioning HingeNet as a robust method for enhancing pre-trained foundation models in downstream beat and downbeat tracking tasks.

\textbf{Powerful Potential of Foundation Models.} Compared to the latest SOTA model Beat This \cite{beatthis}, our proposed method outperforms it across both beat and downbeat accuracy. Specifically, in downbeat F-Measure, our method achieves 79.8\%, surpassing it by as much as 4.3\%. This highlights the significant potential of fine-tuning  music foundation models, which can significantly improve performance in downstream tasks.

Table \ref{other_dataset} compares the performance of HingeNet with several SOTA models on the \textit{Ballroom}, \textit{Hainsworth}, and \textit{SMC} datasets.   Our proposed HingeNet outperforms previous SOTA models in both beat and downbeat tracking across all datasets.

Additionally, we observed that MERT performs better in beat tracking, while MusicFM excels in downbeat tracking. This difference can be attributed to the distinct strengths of each model: MERT's unique music teacher design allows it to capture local musical details well, while MusicFM, with its random projection-based tokenization method, captures long-term contextual dependencies effectively.

\begin{figure}[htbp]
	\begin{minipage}{0.95\linewidth}
		\vspace{0pt}
		\centerline{\includegraphics[width=\textwidth]{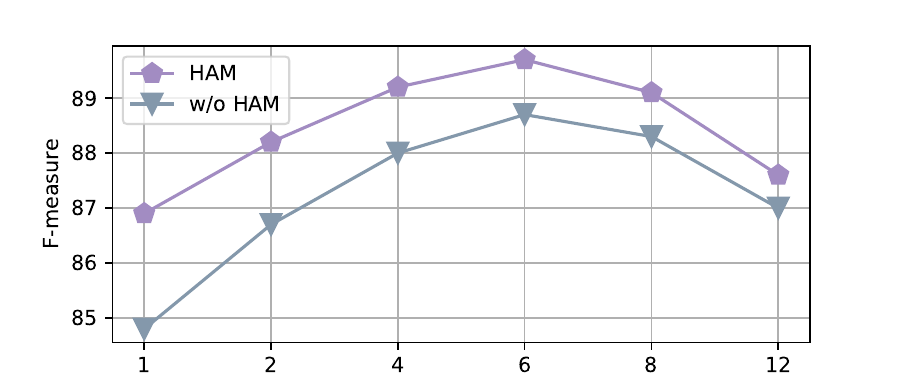}}
	\end{minipage}
	\caption{
  Ablation study on beat tracking with different projection factor $r$ and HAM on the GTZAN dataset.}
	\label{fig3}
\end{figure}

\subsection{Ablation Study}
We conduct two ablation studies to analyze the effectiveness of our proposed method: one focuses on the harmonic-aware module (HAM) and projection factor $r$, and the other evaluates different fine-tuning methods (Adapter, LoRA, and HingeNet).

\textbf{Effect of HAM and Projection Factor.}
From Fig. \ref{fig3}, we can draw three key conclusions: 1) The inclusion of HAM consistently improves performance across all projection factors; 2) The model with HAM achieves the highest beat F-measure when the projection factor $r$ is set to 6; 3) Interestingly, the performance gains from HAM gradually diminish as the projection factor $r$ increases. This diminishing effect is likely due to the distortion of harmonic structures, and this distortion becomes more pronounced as the projection factor $r$ increases, which limits the effectiveness of the HAM in capturing harmonic information. Conversely, with a lower $r$, the feature space remains highly complex, making it challenging to fine-tune the model effectively with limited annotated data. A projection factor of 6 strikes the optimal balance, simplifying the feature space  while preserving the necessary harmonic detail for accurate beat tracking. It is worth mentioning that experiments on other datasets also yielded the best results when the projection factor $r$ is set to 6.

\textbf{Comparison of Fine-Tuning Methods.}
Table \ref{ab_finetune} presents the results of our ablation study, comparing different fine-tuning methods on two foundation models for beat and downbeat tracking. The results clearly demonstrate that HingeNet consistently outperforms other fine-tuning methods. During the experiments, we observed that both Adapter and LoRA exhibited varying degrees of overfitting, which led to their suboptimal performance. In contrast, HingeNet leverages its lightweight and independent structure to fully exploit the capabilities of pre-trained foundation models.

\section{conclusion}
In this paper, we propose HingeNet, a novel and general fine-tuning method specifically designed for beat tracking tasks. HingeNet is a lightweight and separable network with a unique architecture that grants it broad generalizability, enabling effective integration with various pre-trained foundation models.  Furthermore, considering the significance of harmonics in beat tracking, we design the harmonic-aware modules based on harmonic principles to refine and emphasize the harmonic structures in musical signals. Experiments on benchmark datasets demonstrate that HingeNet achieves state-of-the-art performance in both beat and downbeat tracking. 

\section{Acknowledgement}
This work was supported by NSFC(62171138).

\bibliographystyle{IEEEbib}
\bibliography{icme2025references}

\end{document}